\documentstyle[aps,epsf]{revtex}
\newcommand{\tfrac}[2]{{\textstyle\frac{#1}{#2}}}
\begin{document}
\title{Self-Consistent Effective Action for
Quantum Particle with Space-Dependent Mass}

\draft

\author{M. E. S. Borelli\thanks{E-mail: borelli@physik.fu-berlin.de} and
H. Kleinert\thanks{E-mail: kleinert@physik.fu-berlin.de; website:
http://www.physik.fu-berlin/$\sim$kleinert}} \address{Institut
f\"ur Theoretische Physik \\ Freie Universit\"at Berlin \\ Arnimallee 14,
14195 Berlin.}  \date{\today} \maketitle
\begin{abstract}
We calculate the quantum corrections to the classical action of a particle with
coordinate-dependent mass. The result is made self-consistent
by a variational approach, thus making it applicable
to strong-couplings and singular potentials.
By including thermal fluctuations, the we obtain an effective
action  whose classical Euler-Lagrange equation describes
the motion of a particle
including quantum and thermal effects.
\end{abstract}

\section{Introduction}
Recently, Jona-Lasinio's group \cite{cjtp} have calculated the
first-order quantum correction to the classical equation of motion for
a particle in one dimension.  Both the potential and the kinetic
energy terms are modified by terms of order $\hbar$.  In this note, we
extend these results in four essential ways to make them applicable to
nontrivial systems.  First, we allow the mass to be
coordinate-dependent from the outset.  This is necessary to describe a
larger variety of interesting physical systems, for instance compound
nuclei, where the collective Hamiltonian, commonly derived from a
microscopic description via a quantized adiabatic time-dependent
Hartree-Fock theory (ATDHF)\cite{GoRe}, contains coordinate-dependent
collective mass parameters.  Second, we calculate multi-loop
corrections to the effective potential.  Third, we make the result
self-consistent to allow for realistic physical application to systems
with strong-coupling or singular potentials.  Fourth, we include the
effect of thermal fluctuations.

On the technical side of the one-loop calculation, we differ from
\cite{cjtp} by not using the old-fashioned derivative expansion method
of Iliopolous et al. \cite{jim}, since it cannot be generalized to
terms of higher order in derivatives than two, but a more powerful
method of Fraser \cite{caroline}, which can be extended to any
required order.

\section{Quantum corrections to effective action}

\hspace{\parindent} Consider a particle with coordinate-dependent mass
$m(x)$ moving in a one-dimensional potential $V(x)$. Its classical
Lagrangian reads
\begin{equation}
 \label{eq:lagclass}
  {\cal L}(x, {\dot x}) =  \tfrac{1}{2} m(x) {\dot x}^2 - V(x),
\end{equation}
where dots indicate time derivatives.  According to the path integral
formulation of quantum mechanics \cite{feynman,kleinbook}, the
probability amplitude of the particle initially at position $x_a$ at
time $t_a$ to be found at position $x_b$ at a later time $t_b$ is
given by the path integral
\begin{equation}
  \label{eq:amplitude}
  \langle x_a, t_a | x_b, t_b \rangle = \int {\cal D}x
  \exp \left\{ \frac{i}{\hbar} {\cal A}[x] \right\},
\end{equation}
where ${\cal A}[x]$ is the classical action
\begin{equation}
  \label{eq:actionclass}
  {\cal A}[x] = \int_{t_a}^{t_b} {\rm d}t \; {\cal L}(x, {\dot x}),
\end{equation}
and the path integral (\ref{eq:amplitude}) runs over all
paths with fixed end points at $x(t_a)=x_a$ and $x(t_b)=x_b$.

In the path integral formulation, the rules of quantum mechanics
appear as a natural generalization of the rules of classical
statistical mechanics.  In statistical mechanics, each volume in phase
space is occupied with the Boltzmann probability.  In the path
integral formulation of quantum mechanics, each volume element in the
{\em path phase space} is associated with a pure phase factor $\exp \{i
{\cal A}[x]/ \hbar \}$.  One may thus consider the quantum-mechanical partition
function \cite{kleinbook}
\begin{equation}
  \label{eq:quantumpartfunc}
  Z_{\rm QM}(t_b,t_a)[X] \equiv  \int_{\bar x_a = \bar x_b=0} {\cal D}\bar x
  \exp \left\{\frac{i}{\hbar} {\cal A}[X+x]\right\},
\end{equation}
for  the  fluctuations  $\bar x$ around some background
orbit $X(t)$. The result  defines an  effective action in analogy to
the free
energy in quantum statistics:
\begin{equation}
  {\cal A}^{\rm eff} [X]\equiv - i \hbar \ln Z_{\rm QM}[X].
\end{equation}
The Euler-Lagrange equations of motion  extremizing
$
  {\cal A}^{\rm eff} [X]$ account for
 all quantum effects in classical orbits.

Since $Z_{\rm QM}$ cannot in general be
calculated exactly, we must resort to
approximations.  In the
so-called semi-classical approximation,
${\cal A}^{\rm eff}$ is expanded around the classical action, and the quantum
corrections
are expressed as a series expansion in powers of $\hbar$,
also referred to as loop expansion.

Let $x_{\rm cl}(t)$ be the classical path, solving the classical
equation of motion
\begin{equation}
  \label{eq:motionclass}
  m(x_{\rm cl}) \ddot x_{\rm cl} + \tfrac{1}{2} m'(x_{\rm cl}) {\dot
  x_{\rm cl}}^2 +
  V'(x_{\rm cl}) = 0.
\end{equation}
Being an extremum of the action (\ref{eq:actionclass}), the first
functional derivative of ${\cal A}[x]$ vanishes at $x_{\rm cl}(t)$:
\begin{equation}
  \label{funcderiv} \left. \frac{\delta {\cal A}[x]}{\delta x(t)}
  \right|_{x_{\rm cl}}= \frac{\partial {\cal L}}{ \partial x_{\rm cl}}
  - \frac{d}{dt} \frac{\partial {\cal L}}{ \partial \dot x_{\rm cl}} +
  \frac{d^2}{dt^2} \frac{\partial {\cal L}}{\partial \ddot x_{\rm cl}}
  + \cdots = 0.
\end{equation}
Hence ${\cal A}[x]$ has a functional Taylor series starting out like
\begin{equation}
  \label{eq:actionexpansion} {\cal A}[x] = {\cal A}[x_{\rm cl}] +
  \tfrac{1}{2}\left.\frac{\delta^2 {\cal A}[x]}{\delta x(t) \delta
  x(t')}\right|_{x_{\rm cl}} {\bar x}(t){\bar x}(t') + \cdots,
\end{equation}
where ${\bar x}(t) = x(t) - x_{\rm cl}(t)$ and the quantum-mechanical
partition function reads, in the semi-classical approximation,
\begin{equation}
  \label{eq:quantpartfunc_sc} Z_{\rm QM}(t_b,t_a)[x_{\rm cl}] =
  \exp\left\{\frac{i}{\hbar} {\cal A}[x_{\rm cl}]\right\}\int_{x_a =
  x_b}\!\!\!\! {\cal D}{\bar x} \exp\left\{\frac{i}{\hbar} \int_{t_a}^{t_b}
  {\rm d}t \; {\bar x}(t)\left[\left.\frac{\delta^2 {\cal A}[x]}{\delta x(t) \delta
  x(t)}\right|_{x_{\rm cl}} \right] {\bar x}(t) \right\}.
\end{equation}
The path integral in (\ref{eq:quantpartfunc_sc}) is Gaussian, and can
be calculated analytically (for details see Ref.\ \cite{kleinbook}),
yielding an effective action
\begin{equation}
  \label{eq:effecaction_clform}
  {\cal A}^{\rm eff}[x_{\rm cl}]  \equiv {\cal A}[x_{\rm cl}] +  {\cal A}_1[x_{\rm cl}] + {\cal A}^{\rm ml}[x_{\rm cl}] ,
\end{equation}
with the one-loop quantum correction
\begin{equation}
  \label{eq:gamma} {\cal A}_1[x_{\rm cl}] = - \frac{i \hbar}{2} \mbox{Tr} \ln
  \Bigg[m(x_{\rm cl}) {\hat \omega}^2 - V''(x_{\rm cl}) - i m'(x_{\rm
  cl}){\dot x_{\rm cl}}{\hat \omega} - m'(x_{\rm cl}){\ddot x_{\rm
  cl}} - \tfrac{1}{2} m''(x_{\rm cl}) {\dot x_{\rm cl}}^2 \Bigg],
\end{equation}
where we have defined the time derivative operator ${\hat\omega}
\equiv - i {\rm d}/{\rm d}t$. ${\cal A}^{\rm ml}$ contains all multi-loop quantum corrections.
The functional trace Tr in
(\ref{eq:gamma}) contains a time integral $\int_{t_a}^{t_b} dt$ as
well as a discrete sum over all eigenvalues $\omega_n$ of the operator
${\hat\omega}$ \cite{caroline}.  Since the summation over discrete
eigenvalues introduces unnecessary complications in the calculation,
we shall replace the sum $\sum_n f(\omega_n)$ with an integral $\int
({\rm d}\omega / 2 \pi) \; f(\omega)$.  This approximation does not
affect the results as long as the characteristic frequency
along the path
\begin{equation}
 \omega (x)\equiv\sqrt{V''(x_{\rm cl})/m (x)}
\label{fre}\end{equation}
is everywhere much larger than $1/ \Delta t$, where $\Delta t$ is the
inverse time interval where the particle moves very little. The
precise form of this interval will be specified later.

If the classical path $x_{\rm cl}(t)$ is taken to be a constant, the
determination of the functional trace in (\ref{eq:gamma}) is
straightforward, and the quantum correction to the classical
Lagrangian has no explicit time-dependence.  We are interested in
calculating corrections that are explicitly time-dependent, that is,
we look for corrections proportional to $\dot x_{\rm cl}(t)$, $\ddot
x_{\rm cl}(t)$, etc.  The {\em time derivative expansion}
\cite{caroline} of the correction to the effective action has the
general form
\begin{equation}
  \label{eq:effecactio_genform}
  {\cal A}_1[x_{\rm cl}] = \int_{t_a}^{t_b} {\rm d}t \left[ -{\cal V}(x_{\rm
  cl}(t)) + \tfrac{1}{2} {\cal Z}(x_{\rm cl}(t)){\dot x}_{\rm
  cl}^2 + {\cal Z}_2(x_{\rm cl}(t)){\ddot x}_{\rm
  cl}^2 + \cdots \right].
\end{equation}

The idea behind this derivative expansion \cite{caroline} is to
set $x_{\rm cl}(t)$ in (\ref{eq:gamma}) and
(\ref{eq:effecactio_genform}) equal to $x_0 + {\tilde x}(t)$, where
$x_0$ is a constant, and expand both (\ref{eq:gamma}) and
(\ref{eq:effecactio_genform}) in powers of ${\tilde x}$ and its
derivatives. By comparing the result of both expansions, the
coefficients ${\cal V}(x)$, ${\cal Z}(x)$, etc., are extracted.

Let us calculate the first two coefficients in
(\ref{eq:effecactio_genform}).  An expansion around about $x_0$ up to terms
of order ${\tilde x}^2$ yields
\begin{equation}
  \label{eq:firstexpansion}
  {\cal A}_1[x_{\rm cl}] = \int_{t_a}^{t_b} {\rm d}t \left[ -{\cal V}(x_0)  -
  {\cal V}'(x_0){\tilde x} - \tfrac{1}{2} {\cal V}''(x_0){\tilde x}^2
  + \tfrac{1}{2} {\cal Z}(x_0){\dot{\tilde x}}^2 + \cdots \right].
\end{equation}
The corresponding expansion of (\ref{eq:gamma}) yields
\begin{equation}
  \label{eq:secondexpansion}
  {\cal A}_1 = - \frac{i \hbar}{2} \mbox{Tr}
  \ln [G^{-1}({\hat \omega}) + \Lambda(\tilde x)]
  = - \frac{i \hbar}{2} \mbox{Tr} \ln
  [G^{-1}({\hat \omega})] - \frac{i \hbar}{2} \mbox{Tr} \ln
  [1 + G({\hat \omega})\Lambda({\tilde x})],
\end{equation}
where the $G^{-1}({\hat \omega})$ is the inverse free propagator
\begin{equation}
  \label{eq:gm1w}
   G^{-1}({\hat \omega}) = m(x_0) {\hat \omega}^2 - V''(x_0),
\end{equation}
and $\Lambda(\tilde x)$ contains all time dependent terms:
\begin{eqnarray}
  \label{eq:Lambda} \Lambda(\tilde x) &=& {\tilde x}[m'(x_0) {\hat
  \omega}^2 - V'''(x_0)] + \tfrac{1}{2}{\tilde x}^2[m''(x_0) {\hat
  \omega}^2 - V''''(x_0)] - i \dot{\tilde x} m'(x_0) {\hat \omega}
  \nonumber \\ &-& i {\tilde x} \dot{\tilde x} m''(x_0) {\hat \omega}
  - \ddot{\tilde x} m'(x_0) - {\tilde x} \ddot{\tilde x} m''(x_0) -
  \tfrac{1}{2} {\dot {\tilde x}}^2 m''(x_0).
\end{eqnarray}

The first term on the right hand side of (\ref{eq:secondexpansion})
can be calculated immediately: the trace is converted into a simple
integral over time and over the eigenvalues $\omega$ of $\hat \omega$,
since $V''(x_0)$ is time-independent.  Using the integral
formula\cite{gradstein}
\begin{equation}
  \label{eq:intformula}
  \int_{- \infty}^{\infty} \frac{{\rm d} \omega}{2 \pi}
  \frac{\omega^p}{[m \omega^2 - V''(x_0)]^q} = \frac{ i^{-(p+1)} (-1)^{-q}
   [1 + (-1)^p] \Gamma(q - \frac{p+1}{2})
  \Gamma(\frac{p+1}{2})}{4 \pi [V''(x_0)]^{q - \frac{p+1}{2}}
  m^{\frac{p+1}{2}} \Gamma[q]}
\end{equation}
for $p\to 0, q \to 0$ we obtain
\begin{equation}
  \label{eq:effecpot}
  {\cal V}(x_0) = \frac{\hbar}{2} \sqrt{\frac{V''(x_0)}{m(x_0)}}.
\end{equation}
Replacing $x_0$ by $x_{\rm cl}(t)$,
this determines the lowest-order
term in the derivative expansion (\ref{eq:effecactio_genform}), in
agreement with Ref.\ \cite{cjtp}.  The logarithm in the second term in
(\ref{eq:secondexpansion}) is now expanded up to order ${\tilde x}^2$:
\begin{equation}
  \label{eq:logexpansion}  - \frac{i \hbar}{2} \mbox{Tr }\ln [1 +
  G({\hat \omega})\Lambda(x)] =  - \frac{i \hbar}{2} \mbox{Tr}
  [G({\hat \omega})\Lambda({\tilde x})] + \frac{i \hbar}{4} \mbox{Tr}
  [G({\hat \omega})\Lambda({\tilde x})G({\hat \omega})\Lambda({\tilde
  x})] + \cdots .
\end{equation}
The first term on the right hand side is calculated using formula
(\ref{eq:intformula}),
\begin{eqnarray}
  \label{eq:firstterm} \frac{i \hbar}{2} \mbox{Tr} [ G({\hat
  \omega})\Lambda({\tilde x})] = \int_{t_a}^{t_b} {\rm d}t \;
  \frac{\hbar}{4} \Bigg\{ && \Bigg[\frac{m'(x_0)}{[m(x_0)]^{3/2}}
  \sqrt{V''(x_0)}- \frac{V'''(x_0)}{\sqrt{m(x_0) V''(x_0)}}\Bigg]
  {\tilde x} \nonumber \\ && +\frac{1}{2}\Bigg[
  \frac{m''(x_0)}{[m(x_0)]^{3/2}}
  \sqrt{V''(x_0)}-\frac{V''''(x_0)}{\sqrt{m(x_0) V''(x_0)}}
  \Bigg]{\tilde x}^2 \nonumber \\ &&
  +\frac{1}{2}\frac{m''(x_0)}{\sqrt{m(x_0)V''(x_0)}}{\dot{\tilde
  x}}^2 + \cdots \Bigg\} .
\end{eqnarray}
By comparing the linear term in ${\tilde x}$ with that in
(\ref{eq:firstexpansion}) we identify
\begin{equation}
  \label{eq:checkfirstderivative}
  {\cal V}'(x_0) = - \frac{\hbar}{4} \Bigg[\frac{m'(x_0)}{[m(x_0)]^{3/2}}
  \sqrt{V''(x_0)}- \frac{V'''(x_0)}{\sqrt{m(x_0) V''(x_0)}}\Bigg].
\end{equation}
To find the quadratic terms in (\ref{eq:logexpansion}), we must
first move all operators ${\hat \omega}$ to the left, and all
functions of $t$ to the right.  Then we can perform the traces
independently.  To do this, we formulate the commutator
  $[f(t),{\hat \omega}] = i {\dot f}(t)$
applied as a  rule
\begin{equation}
  \label{eq:identity}
  f(t){\hat \omega} g(t) = \left[{\hat \omega} + i \partial_t \right]f(t)g(t),
\end{equation}
with the convention that
time derivative operators $\partial_t$ act {\em only} on the first
function on its right.  By repeatedly applying this rule, the
second term in (\ref{eq:logexpansion}) may be expanded in powers of
time derivatives of $\tilde x$, and gives, up to ${\cal O}(\dot{\tilde
x})^2$
\begin{eqnarray}
  \label{eq:derivexpansion}
  \mbox{Tr}[G({\hat \omega})\Lambda({\tilde x}) G({\hat \omega})
   \Lambda({\tilde x})] &=& \mbox{Tr} [G^2({\hat
  \omega})\Lambda({\tilde x})\Lambda({\tilde x})] + \mbox{Tr}[
   G^3({\hat \omega}) \left(- 2 i m \omega \partial_t + m
   \partial_t^2 \right) \Lambda({\tilde x})\Lambda({\tilde x})]
   \nonumber \\ &+& \mbox{Tr} [ G^4({\hat \omega}) \left( - 2 i m
   \omega \partial_t + m \partial_t^2 \right)^2 \Lambda({\tilde
   x})\Lambda({\tilde x})].
\end{eqnarray}
In each line, the derivative operators $\partial_t$ act only on the
first $\Lambda(\tilde x)$.  After carrying out all integrations, we
are left with
\begin{eqnarray}
  \label{eq:secondterm} \lefteqn{ \frac{i \hbar}{2} \mbox{Tr}
  \left[G({\hat \omega})\Lambda({\tilde x})G({\hat
  \omega})\Lambda({\tilde x}) \right] =} && \nonumber \\ &&
  \;\;\;\;\;\;\;\;\;\;\;\;\;\;\int_{t_a}^{t_b}
  {\rm d}t \Biggl\{\left[-\frac{3\hbar}{16} \frac{[m'(x_0)]^2
  [V''(x_0)]^{1/2}}{[m(x_0)]^{5/2}} +\frac{\hbar}{8} \frac{m'(x_0)
  V'''(x_0)}{[m(x_0)]^{3/2}[V''(x_0)]^{1/2}} + \frac{\hbar}{16}
  \frac{[V'''(x_0)]^2}{[m(x_0)]^{1/2}[V''(x_0)]^{3/2}}\right]{\tilde
  x}^2 \nonumber \\ && \;\;\;\;\;\;\;\;\;\;\;\;\;\;\;\;\;\;\;\;\;\;\;\;+ \left[\frac{\hbar}{64} \frac{[V'''(x_0)]^2
  [m(x_0)]^{1/2}}{[V''(x_0)]^{5/2}} -\frac{5 \hbar}{32} \frac{m'(x_0)
  V'''(x_0)}{[m(x_0)]^{1/2}[V''(x_0)]^{3/2}} - \frac{11 \hbar}{64}
  \frac{[m'(x_0)]^2}{[m(x_0)]^{3/2}[V''(x_0)]^{1/2}}\right]{\dot{\tilde
  x}}^2 \Biggr\}. \label{trace_result}
\end{eqnarray}

Adding the second term in (\ref{eq:firstterm}) to the coefficient of
the ${\tilde x}^2$ term in the above expression we obtain $-{\cal
V}''(x_0)/2$, as necessary for the consistency of the expansion
(\ref{trace_result}) with (\ref{eq:effecactio_genform}).  From the
second term in (\ref{eq:secondterm}), proportional to ${\dot {\tilde
x}}^2$, combined to the correspondent term in (\ref{eq:firstterm}), we
extract an $x_0$-dependent contribution to the kinetic energy
(\ref{eq:effecactio_genform}):
\begin{eqnarray}
 \label{eq:zterm}
 && {\cal Z}(x_0)= \frac{\hbar}{32}
  \frac{[V'''(x_0)]^2 [m(x_0)]^{1/2}}{[V''(x_0)]^{5/2}} -\frac{5 \hbar}{16}
  \frac{m'(x_0) V'''(x_0)}{[m(x_0)]^{1/2}[V''(x_0)]^{3/2}} \nonumber
   - \frac{11 \hbar}{32}
  \frac{[m'(x_0)]^2}{[m(x_0)]^{3/2}[V''(x_0)]^{1/2}} + \frac{\hbar}{4}
  \frac{m''(x_0)}{[m(x_0)]^{1/2}[V''(x_0)]^{1/2}}. \nonumber \\ &&
\end{eqnarray}
We now include all local multi-loop diagrams
in the well-known way  \cite{feynklein,kleinbook}.
These smears out the anharmonic part of the potential
$V^{\rm int}(x)\equiv V(x)-V''(x)x^2/2$ to
\begin{equation}
  \label{eq:va2} V^{\rm int}_{a^2}(x) = \int \frac{{\rm d}
x'}{\sqrt{2 \pi a^2(x)}} \exp \left[ - \frac{(x - x')^2}{2
a^2(x)}\right] V^{\rm int}(x')
\end{equation}
 over the harmonic fluctuation width
\begin{equation}
 \label{eq:a2} a^2= a^2_\omega(x) = \langle {\bar
x}^2(t)\rangle^\omega = \frac{\hbar}{2M \omega (x)}.
\end{equation}
We observe that, for slow enough particle movement, the time interval
$\Delta t$ determining the range of validity of our calculations is
given by
\begin{equation}
 \Delta t = \frac{m(x) a^2_\omega(x)}{\hbar}.
\end{equation}
Replacing $x_0$ everywhere by the position of the background path $X$,
the total effective action becomes
\begin{equation}
  \label{eq:effecaction_final} {\cal A}^{\rm eff}[X] = \int_{t_a}^{t_b} {\rm
  d}t \;\; \left[\tfrac{1}{2} m^{\rm eff}(X) {\dot X}^2 -
V^{\rm eff}_{a_\omega^2}(X)\right],
\label{our}\end{equation}
where the effective potential is given by
\begin{equation}
  \label{eq:veff}
   V^{\rm eff}[X] = \tfrac{1}{2} \hbar \omega(X) +
   V^{\rm int}_{a_\omega^2}(X).
\end{equation}
calculated within a harmonic path integral containing a trial
frequency $\Omega$ to be determined by minimization of
$V_{a^2_\omega}^{\rm int}$ (for details, see Ref.\ \cite{kleinbook}).
The effective mass appearing in Eq.\ (\ref{eq:effecaction_final}) is
from (\ref{eq:zterm})
\begin{eqnarray}
  \label{eq:meff}
   m^{\rm eff}_ \omega (X) = m(X) + \frac{\hbar}{16}
  \frac{[\omega'(X)]^2}{\omega^3(X)} -\frac{\hbar}{4}
  \frac{\omega'(X)}{\omega(X)} \frac{m'(X)}{m(X)}
  - \frac{5 \hbar}{16}
  \frac{1}{\omega(X)}\frac{[m'(X)]^2}{m^2(X)} + \frac{\hbar}{8}
  \frac{1}{\omega(X)}\frac{m''(X)}{m(X)}.
\end{eqnarray}
If we omit
the multi-loop corrections, and the $X$-dependence of the mass,
our effective potential (\ref{our}) reduces to the expression
found by \cite{cjtp}.

Our  result is now made self-consistent as in variational
perturbation theory in \cite{feynklein,kleinbook}.  We simply replace
$\omega(X)$ by a trial frequency $\Omega(X)$ and add to the
action the term
\begin{equation}
  \Delta {\cal A}^{\rm eff} = - \frac{m(X)}{2} \left[ \omega^2(X) -
  \Omega^2(X)\right] a^2_{\Omega}(X),
\end{equation}
which is the expectation value of the difference between the original
and the variational potential.  For a slowly moving particle, the
combined action may be extremized in $\Omega(X)$ for a constant
background ${\dot X}(t)=0$.  This yields the self-consistency equation
\begin{equation}
 \Omega ^2(X)=
\frac{2}{m(X)}\frac{\partial }{\partial _{a^2}}
V_{a^2}(X)=\frac{1}{m(X)}\left[ \frac{\partial ^2}{\partial x_0^2}V_{a^2 }
\right] _{a^2=a_ \Omega ^2(X)}.
\label{at}\end{equation}
  This is inserted in $m^{\rm eff}_ \Omega (X)$ to
obtain the fluctuation-corrected mass term.

Finally we may include also thermal fluctuations.
If the particle moves slowly  enough  to allow for an approximate thermal equilibration
at each point of  the orbit,
we may simply replace the quantum mechanical fluctuation width
(\ref{eq:a2}) at $X$ and $ \Omega (X)$ by the quantum statistical
\cite{feynklein,kleinbook}:
\begin{equation}
a^2_ \Omega (X)=\frac{k_{\rm B} T}{m(X) \Omega ^2(X)}\left[\frac{
\Omega (X)}{2 k_{\rm B} T}\coth \frac{\Omega (X)}{2 k_{\rm B} T} -1\right] ,
\label{at2}\end{equation}
where $k_{\rm B} T$ is the thermal energy.

The resulting effective action gives rise to a new {\em classical}
equation of motion, which should help improving our understanding of
the relation between classical and quantum physics.

\end{document}